\documentclass[aps,prx,twocolumn,nopacs,superscriptaddress]{revtex4}

\usepackage{graphicx}
\usepackage{verbatim}
\usepackage{amsmath}
\usepackage{sidecap}
\usepackage{color,soul}
\usepackage[nottoc]{tocbibind}

\begin{document}

\title{Experimental signatures of phase interference and sub-femtosecond time dynamics on the incident energy axis of resonant inelastic X-ray scattering}

\author{L. Andrew Wray}
\email{lawray@nyu.edu}
\thanks{Corresponding author}
\affiliation{Department of Physics, New York University, New York, NY 10003, USA}
\author{Shih-Wen Huang}
\affiliation{Advanced Light Source, Lawrence Berkeley National Laboratory, Berkeley, CA 94720, USA}
\author{Yuqi Xia}
\author{M. Zahid Hasan}
\affiliation{Department of Physics, Joseph Henry Laboratories, Princeton University, Princeton, NJ 08544, USA}
\author{Charles Mathy}
\affiliation{ITAMP, Harvard-Smithsonian Center for Astrophysics, Cambridge, MA 02138, USA}
\author{Hiroshi Eisaki}
\affiliation{Nanoelectronic Research Institute, National Institute of Advanced Industrial Science and Technology, Tsukuba, 305-8568, Japan}
\author{Zahid Hussain}
\author{Yi-De Chuang}
\affiliation{Advanced Light Source, Lawrence Berkeley National Laboratory, Berkeley, CA 94720, USA}

\begin{abstract}

Core hole resonance is used in X-ray spectroscopy to incisively probe the local electronic states of many-body systems. Here, resonant inelastic X-ray scattering (RIXS) is studied as a function of incident photon energy on Mott insulators SrCuO$_2$ and NiO to examine how resonance states decay into different excitation symmetries at the transition metal M-, L- and K-edges. Quantum interference patterns characteristic of the two major RIXS mechanisms are identified within the data, and used to distinguish the attosecond scale scattering dynamics by which fundamental excitations of a many-body system are created. A function is proposed to experimentally evaluate whether a particular excitation has constructive or destructive interference in the RIXS cross-section, and corroborates other evidence that an anomalous excitation is present at the leading edge of the Mott gap in quasi-one dimensional SrCuO$_2$.

\end{abstract}


\date{\today}

\maketitle


\section{Introduction}

Dynamics of some of the fastest electronic processes in nature, including the motion of electrons between neighboring atomic orbitals, are encoded in the combined frequency and phase profile of resonant photon scattering experiments \cite{TohyamaSqw,DevereauxSqw,VDBultrashort,AmentRIXSReview,WrayNiO,WrayCoO,NCCOIshii,trXASsim}. However, current frequency-resolved resonant scattering measurements cannot be converted to a time resolved picture, because techniques that determine relative phase information from resonant elastic scattering have not been adapted to the greater complexity of inelastic spectra. In resonant inelastic X-ray scattering (RIXS), core hole resonance modes are used to enhance coupling between photons and low energy electronic degrees of freedom, resulting in inelastic spectral features that typically include more than one overlapping excitation symmetry. These final states of RIXS include modes that are of broad fundamental interest, such as momentum-dependent Mott gap transitions and emergent particles composed of diverse combinations of charge, orbital and spin degrees of freedom \cite{Hasan_KRIXS,HasanSCO_KRIXS,TohyamaSqw,DevereauxSqw,YJKimSCO,NCCOIshii,NCCOyinwan,WrayNiO,WrayCoO,VDBultrashort,AmentRIXSReview,Jeroen1Dfluct,spinOrbitalSep,NiO_LRIXS,NiO_MRIXS,KotaniSIAM,FC_K,FC_L,magModes1,magModes2,magModes3,magModes4,Valentina}.

Within the Kramers-Heisenberg scattering equation, phase information is reflected in interference patterns that occur when multiple quantum paths of resonance lead to the same final state. The quantum paths considered in this study are identified with respect to the incident photon energies they resonate with, rather than spatial coordinates. Measuring the relative phases associated with different resonance paths will provide a new basis for understanding the time dynamics of resonant X-ray scattering, due to the conjugate nature of time and energy, and will give a new dimension of information by which to understand the symmetries of RIXS excitations. When the path-phases leading to a given final state interfere constructively, it indicates that this final state is a quickly accessible excitation with quantum symmetries closely related to the core hole resonance chosen for measurement (e.g. $M$, $L_3$). On the other hand, destructive path-phase interference occurs when a final state is accessed through a relatively time-consuming mechanism such as core hole shake-up. This fundamental dichotomy of scattering processes is discussed in Ref. \cite{AmentRIXSReview}, although later literature has not followed identical terminology conventions \cite{MLdirectNote}.

Here, we report the measurement of quantum phase interference patterns in the incident energy dependence of sharply resolved RIXS spectra of Mott insulators ($\delta E \lesssim 35meV$ at the M-edge, and $\delta E \sim 100meV$ at the K-edge). Two-peak interference patterns in the RIXS spectra of a cuprate (SrCuO$_2$) are measured to establish the distinct line shapes of constructively and destructively interfering resonance channels. A basic metric is then introduced to estimate the relative phase information in more complex spectra at the M-, L- and K- resonance edges. This metric provides a new dimension of spectroscopic information that is used to confirm the presence of an anomalous excitation close to the Mott gap of SrCuO$_2$ in K-edge data. More generally, identifying quantum interference is shown to reveal essential features of the sub-femtosecond time dynamics by which excitations are created in resonant interactions between X-rays and matter.

\section{Measurement of simple quantum interference patterns}

\subsection{Quantum paths in the scattering equation}

Energy-resolved resonant scattering is described using the Kramers-Heisenberg equation:

\begin{align} \label{eq:KH}
    R_{f}(E,h\nu) \propto \sum_{g}\left|\sum_m\frac{
    \langle f|T^{\dagger}| m\rangle
    \langle m|T|g\rangle}{h\nu-E_m+i\Gamma_m/2}
    \right|^2 \nonumber \\
    \times\frac{\frac{1}{2\pi}\Gamma_{f}}{(E-E_{f})^2+(\frac{1}{2}\Gamma_{f})^2}
\end{align}

The quantity inside absolute value brackets represents the matrix elements for quantum paths from the ground state $|g\rangle$ to a final state $|f\rangle$, through short-lived intermediate resonance states ($|m$$\rangle$) that have inverse lifetimes $\Gamma_m$ and resonance energies E$_m$. Summation over $|g\rangle$ is included to consider all degenerate atomic ground states (for example, NiO has 12 degenerate spin domains). The term path-phase refers to the complex phase of the numerator ($\langle f|T^{\dagger}| m\rangle \langle m|T|g\rangle$). Incident photon energy is written as $h\nu$, the excitation energy is $E$, and the inverse final state excitation lifetime is $\Gamma_{f}$. The incident photon operator $T$ couples to the polarization of the incident photon beam, while the outgoing photon operator $T^{\dagger}$ couples to the resultant polarization of scattered photons.

The form of Kramers-Heisenberg has several immediate consequences for quantum interference \cite{AmentRIXSReview,MLdirectNote}. In particular, there is a natural division between excitations that can be created purely from the dipole matrix elements of the incoming and outgoing photons (i.e. $\langle f|T^{\dagger}T|g\rangle \neq 0$), and those that can only be created if intermediate states are not summed completely coherently (i.e. $\sum_m\langle f|T^{\dagger}| m\rangle \langle m|T|g\rangle = 0$, but $\sum_m\left|\langle f|T^{\dagger}| m\rangle \langle m|T|g\rangle \right|^2 \neq 0$). The former case, which we will refer to as ``photon operator" RIXS, describes a range of orbital excitations at the transition metal L- and M-edges, and can be considered to apply to single magnon excitations at $L_3$- or $L_2$- edges provided that the edges are well separated from one another.

The latter case, which we will call ``shake-up" RIXS applies to the non-pre-edge features at the transition metal K-edge \cite{Hasan_KRIXS,TohyamaSqw,DevereauxSqw,NCCOIshii,AmentRIXSReview}, and includes principle scattering channels for double magnon (see discussion in Ref. \cite{Valentina}) and charge transfer excitations at the M- and L- edges.  Because $\sum_m\langle f|T^{\dagger}| m\rangle \langle m|T|g\rangle = 0$ for shake-up RIXS, these excitations must have fully destructive path phase interference, and their fractional contribution to scattering intensity in the RIXS spectrum scales as $I\sim K^2/\Gamma^2$ when core hole lifetime is short ($|K/\Gamma| \ll 1$). Here, K is a factor determined by the kinetics through which the resonance states evolve to create a particular excitation, and $\Gamma$ is the intermediate state lifetime. For longer core hole lifetime or faster excitation kinetics ($|K/\Gamma| \gtrsim 1$), the fractional intensity of shake-up excitations converges on a fixed value and phase interference goes from destructive to neutral in Eq. \ref{eq:KH}.

\subsection{2-slit-like quantum interference in a $d^9$ cuprate}

\begin{figure}[t]
\includegraphics[width = 8.7cm]{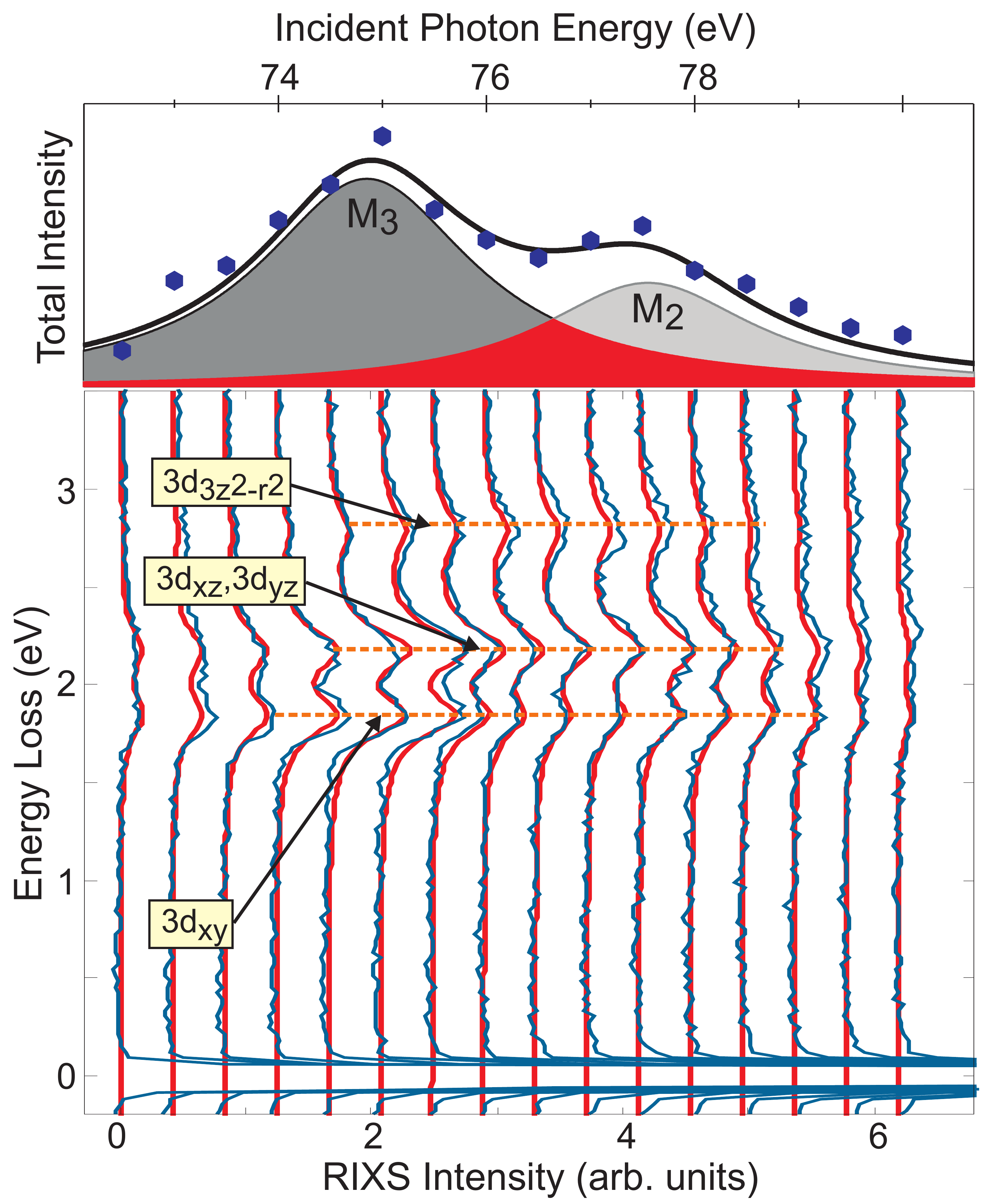}
\caption{{\bf{Overlapping core hole resonance states}}: (top) Inelastic fluorescence yield at the Cu M-edge of SrCuO$_2$ is fitted by two overlapping Lorentzians with a 2:1 intensity ratio. (bottom) (blue) RIXS spectra measured at incident energies corresponding to the x-axis of the top panel reveal three energy-loss peaks representing d-d excitations. Red curves show an atomic multiplet simulation.}
\label{fig:SCOdata}
\end{figure}

\begin{figure}[t]
\includegraphics[width = 8.7cm]{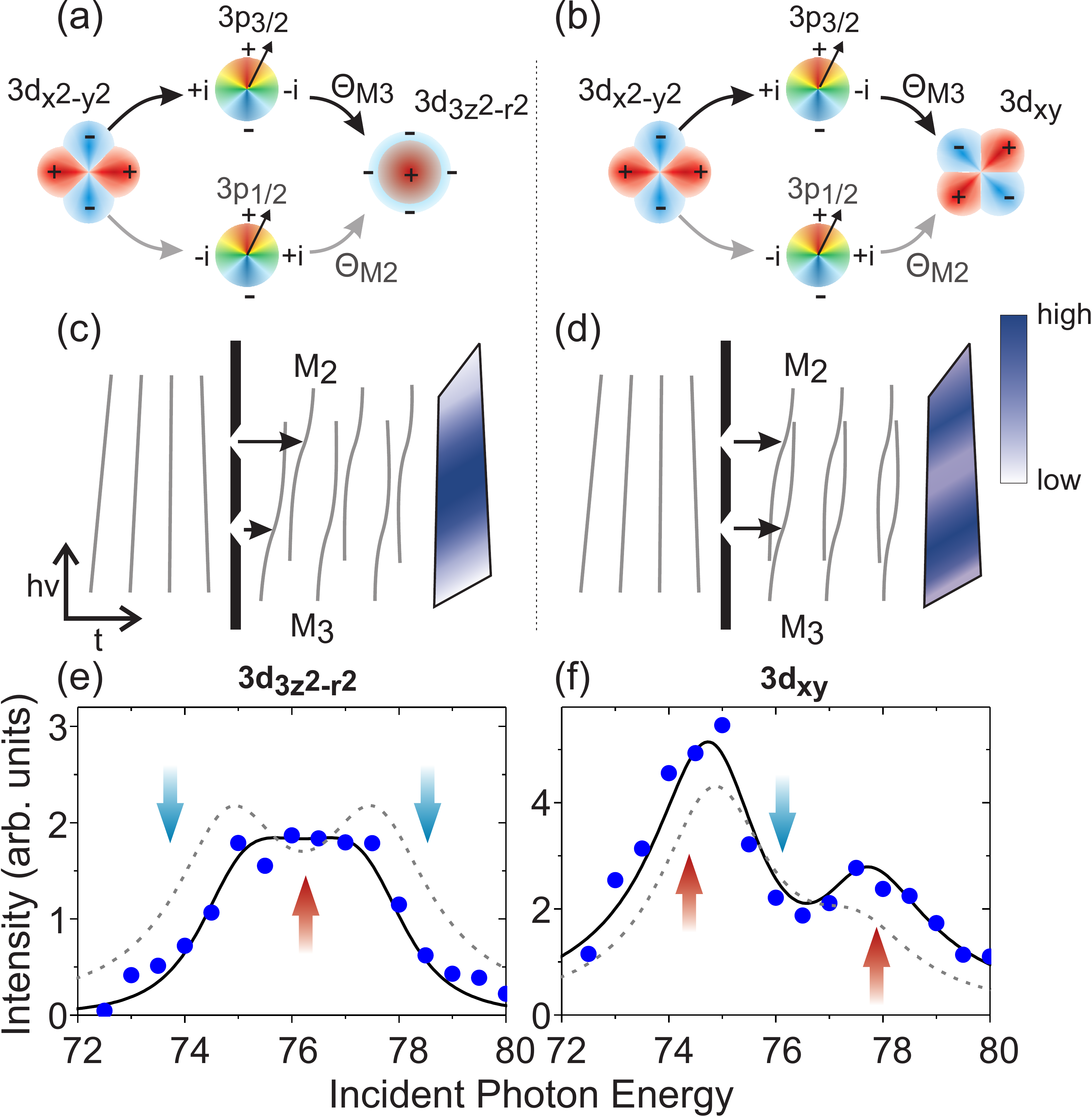}
\caption{{\bf{2-slit RIXS quantum interference patterns}}: The orbital transitions involved in creating a (a) $3d_{3z^2-r^2}$ or (b) $3d_{xy}$ excitation through $M_3$ and $M_2$ core hole resonances are shown. (gray lines) Constant phase contours are shown for the scattering process in which (c) $3d_{3z^2-r^2}$ and (d) $3d_{xy}$ orbiton excitations are created in SrCuO$_2$. The constant phase contours from scattering through $M_3$ and $M_2$ resonance are offset in time by (black arrows) the phase difference between $\Theta_{M2}$ and $\Theta_{M3}$, and interfere constructively when they converge. (e-f), The resulting interference patterns are seen from (circles) incident energy dependence of the $3d_{3z^2-r^2}$ and $3d_{xy}$ RIXS intensity, and (solid line) fitted using Equation (\ref{eq:QIsimpleFit}) with phase shifts $\theta_{M3,z^2}=\theta_{M2,z^2}+\pi$ and $\theta_{M3,xy}=\theta_{M2,xy}$. A dashed line shows the fit result with quantum interference disregarded by moving the sum over intermediate states ($\sum_m$) outside of the absolute value brackets in Equations (\ref{eq:KH},\ref{eq:QIsimpleFit}). Arrows highlight the change from quantum interference.}
\label{fig:SCO2slit}
\end{figure}

The M-edge of cuprates provides several scattering scenarios in which the scattering equation can be rewritten as a 2-channel process, giving a simple and experimentally identifiable ``2-slit" interference pattern.  In this section, we will measure interference patterns that are characteristic of photon operator RIXS and shake-up RIXS in scattering from SrCuO$_2$. The interference patterns will be fitted to estimate underlying phase information for the scattering process and discuss the kinetics by which excitations are created.

The ground state of SrCuO$_2$ has a single hole in the $3d_{x^2-y^2}$ orbital of each copper atom \cite{ZXOrbitals}. Intermediate states of M- and L-edge resonant scattering have full $3d^{10}$ orbital occupation. With no vacant valence orbitals, an atomic multiplet (AM) picture has only two types of intermediate states determined by total angular momentum J=3/2 ($M_3$) and J=1/2 ($M_2$) spin-orbit symmetries. Measurements of RIXS orbital excitations in SrCuO$_2$ across a wide range of incident photon energies are presented in Fig. \ref{fig:SCOdata}(bottom), and have been overlaid with a multiplet simulation (red curves). The visible RIXS features represent excited states in which the hole has been moved to a different orbital, and behave predominantly as orbitons \cite{spinOrbitalSep} due to the lack of strong localization in this quasi-one dimensional material \cite{HasanSCO_KRIXS,spinChargeSep}. These orbitons are labeled by the d-orbital symmetry of the hole. 

Summing RIXS intensity in the 1-3.5eV energy loss range gives the total probability that an orbiton of any symmetry will be excited, which we plot as a function of incident energy in Fig. \ref{fig:SCOdata}(top). In this plot, humps centering on the M$_3$ and M$_2$ resonance energies are well fitted using two closely spaced Lorentzian functions with a 2:1 intensity ratio derived from the core level degeneracies. In contrast, inspection shows that the intensity of individual excitation modes has distinctive dependence on the incident photon energy: the 3d$_{3z^2-r^2}$ symmetry orbiton primarily resonates around a single energy of h$\nu$=76eV, while the 3d$_{xy}$ orbiton resonates over a much broader energy range with two distinct peaks approximately 3eV apart (see Fig. \ref{fig:SCO2slit}(e-f)). The 3eV energy difference is larger than the separation between 3p$_{1/2}$ and 3p$_{3/2}$ core levels. The incident energy dependence for 3d$_{xz}$ and 3d$_{yz}$ modes is not plotted, as they cannot be resolved from one another on the energy loss axis.

The considerable overlap between $M_3$ and $M_2$ resonances seen in Fig. \ref{fig:SCOdata}(top) suggests that orbitons may be excited via intermediate states in which the core hole occupies a coherent quantum superposition of the $M_3$ and $M_2$ levels. To focus on quantum interference between $M_3$ and $M_2$ scattering channels, it is convenient to rewrite the Kramers-Heisenberg equation as:
\begin{align} \label{eq:QIsimpleFit}
    R_{f}(h\nu) \propto \sum_{g}\left|\sum_m A_{f,g,m}e^{i\Theta_{f,g,m}}G_m(h\nu)\right|^2 \nonumber \\
\end{align}
Here, the numerator of Equation (\ref{eq:KH}) is converted into a real valued amplitude $A_{f,g,m}$ and a path-phase of $\Theta_{f,g,m}$. The denominator in Equation (\ref{eq:KH}) is represented by the Green's function $G_m(h\nu)=(h\nu-E_m+i\Gamma_m/2)^{-1}$, and the energy loss (E) axis has been neglected as it is simply a normalized Lorentzian.

The incident energy dependence of excitations described by Equation (\ref{eq:QIsimpleFit}) allows a close analogy to be drawn with 2-slit interference (i.e. Young's experiment). The interference between constant phase contours of wave packets scattered through $M_3$ and $M_2$ core levels can be seen in dimensions of energy and time (Fig. \ref{fig:SCO2slit}(c-d)), rather than the spatial dimensions of normal 2-slit interference. In a standard 2-slit scattering diagram, constant phase contours represent points at which light that has traversed a specified aperture carries a given phase. In the case of RIXS, constant phase contours after the scattering event in Fig. \ref{fig:SCO2slit}(c-d) refer to points at which the phase of the final state wavefunction has a given value, with the system having evolved through a specified resonance state. Time evolution of the phase $\Theta(t)$ is described in the usual way as $\Psi_f(t,h\nu) = e^{-iHt}\Psi_f(0,h\nu) = e^{i\Theta(t)}\Psi_f(0,h\nu)$. Here, $H$ is the Hamiltonian and $\Psi_f(0,h\nu)$ is a final eigenstate that includes a RIXS excitation and a scattered photon, and thus has energy equal to the incident photon energy.

Constant phase contours after scattering adopt a twice-bent shape derived from the phase component of the Green's function $G_m(h\nu)$, and are drawn to the right of the slits in the Fig. \ref{fig:SCO2slit}(c-d) diagrams. The phase component of $G_m(h\nu)$ is symmetric about the $M_3$ or $M_2$ ``slit" resonance energy indexed by `m', and has a maximal rate of change $\left|\frac{\partial \Theta(t)}{\partial h\nu}\right|$ at the resonance energy. The excitation pattern depends on the energy separation between the $M_3$ and $M_2$ resonances, just as the interference pattern in real-space two slit interference is influenced by the spatial separation between the two slits. The difference between $M_3$ and $M_2$ path-phases is indicated with black arrows in Fig. \ref{fig:SCO2slit}(c-d). For the 3d$_{xy}$ and 3d$_{3z^2-r^2}$ orbitons, photon matrix elements allow the f and g indices to be neglected (i.e. $A_{f,g,m}\rightarrow A_{xy,m}$ or $A_{z^2,m}$), by approximating that the 3d$_{xy}$ (3d$_{3z^2-r^2}$) state is always excited without (with) a spin flip \cite{SalaLedge,Kuiper}. These approximations account for more than $90\%$ of scattering events, which is adequate given the intensity jitter on the incident energy axis of our data. A more detailed discussion of spin-related scattering matrix elements is presented at the end of Appendix B.

Fitting the fluoresence peaks in Fig. \ref{fig:SCOdata}(top) gives values of E$_{M3}$=74.85eV, E$_{M2}$=77.55eV, with error similar to $\pm$0.1eV. Core hole lifetime is fitted with just one value of $\Gamma_m$=2.5eV for all intermediate states, because in AM calculations, all 3p core hole states of cuprates have the same decay rate. The peak intensities of individual excitations as a function of incident energy are plotted in Fig. \ref{fig:SCO2slit}(e-f), and are fitted with amplitudes of A$_{M3,xy}$=2.5, A$_{M2,xy}$=1.4, A$_{M3,3z^2-r^2}$=A$_{M2,3z^2-r^2}$=1.7. Incident energy dependence conforms with the phase relation $\Theta_{M2,xy}-\Theta_{M3,xy}=0$ for the 3d$_{xy}$ orbiton, which has the effect of spreading the 3d$_{xy}$ resonance peaks apart. To describe the data, destructive interference between path-phases ($|\Theta_{M2,z^2}-\Theta_{M3,z^2}|\sim\pi$) is required to pinch resonant intensity towards a central energy as seen in the data. In this way the full matrix elements for $M_3$ and $M_2$ paths are determined, except for a global multiplicative constant that can be ignored if scattering intensity is plotted with arbitrary units.

This analysis illustrates that phase information can be retrieved from inelastic X-ray scattering data, and identifies interference patterns characteristic of constructive and destructive path phase interference in a simple 2-resonance case. The phase shifts for this simple case can also be calculated exactly by using the multiplet model (transitions are illustrated in Fig. \ref{fig:SCO2slit}(a-b)), and the 0 and $\pi$ phase shifts from calculations match the results obtained from fitting the resonance profiles. The destructive phase shift in creating 3d$_{3z^2-r^2}$ orbitons can be understood because the 3d$_{3z^2-r^2}$ mode occurs with a spin flip, and spin flips are only allowed via shake-up RIXS at the M- edge. (i.e. $\sum_m\langle f|T^{\dagger}| m\rangle \langle m|T|g\rangle = 0$ for spinful excitations, if m indexes all 3p or 2p core hole states and 3d spin orbit coupling is neglected) As noted in Section 2A, shake-up RIXS excitations must have fully destructive path phase interference, which implies a phase difference of $\pi$ when there are only two resonance channels ($M_2$ and $M_3$). 

The ``shake-up" or ``photon operator" nature of path-phase interference associated with different excitation types at the M-, L- and K-edges will be discussed further in the next section, and a summary table is included in Appendix E. Recognizing phase information in experimental spectra will provide a model-independent metric to distinguish between a wide variety of degenerate excitations that differ in whether they are associated with fast or slow excitation processes. For example, this is a principle distinction between single- and multi-paramagnon features \cite{Valentina}, and between different species of charge excitation as discussed in Sections 3B-C below.

\section{Quantum interference in complex inelastic spectra}

\subsection{Defining a metric}

The results thusfar take advantage of the particularly simple ``2-slit" M-edge resonance scenario in cuprates to establish that full patterns of constructive and destructive quantum phase interference can be distinguished in the incident energy dependence of well-resolved RIXS spectra. Extending the analysis to more complex spectra requires a new approach, as the standard Kramers-Kronig method that is used to estimate how the phase angle varies with incident energy in elastic spectroscopies \cite{KKref} cannot be adapted directly to resonant inelastic spectra without extremely accurate data, absorption normalization and \emph{a priori} theoretical knowledge of the inelastic excitation symmetries. (For phase analysis on the energy loss and momentum axes of \emph{non-resonant} inelastic scattering, see Ref. \cite{Abbamonte_IXS}.) Nevertheless, a metric can be derived to estimate the degree of quantum interference in complex RIXS spectra by focusing on the leading edge of resonance, where constructive path-phases yield softer line shapes in the Kramers-Heisenberg formalism (derivation in Appendix C). We propose one such metric as follows:

\begin{align} \label{eq:zetaDef}
\zeta(E,h\nu)=I(h\nu)^{-1} \frac{\mathrm{d} I(h\nu)}{\mathrm{d} h\nu}-R(E,h\nu)^{-1} \frac{\mathrm{d} R(E,h\nu)}{\mathrm{d} h\nu},
\end{align}

where $R(E,h\nu)$ is the RIXS intensity at incident energy $h\nu$ and energy loss $E$, and the resonant part of X-ray absorbtion (XAS) or fluorescence yield (FY) ($I(h\nu)$) is used to center the function on zero. Good statistics are obtained by evaluating $\zeta(E,h\nu)$ at an incident energy that has maximal slope in the XAS profile. Evaluating at slightly lower incident energies will reduce error from resonant self-absorption, which can be significant in the X-ray regime (L- and K-edges), though it is more negligible at the M-edge \cite{WrayNiO,WrayCoO}.

\subsection{The M-edge of NiO}

\begin{figure*}
\centering
\includegraphics[width = 15cm]{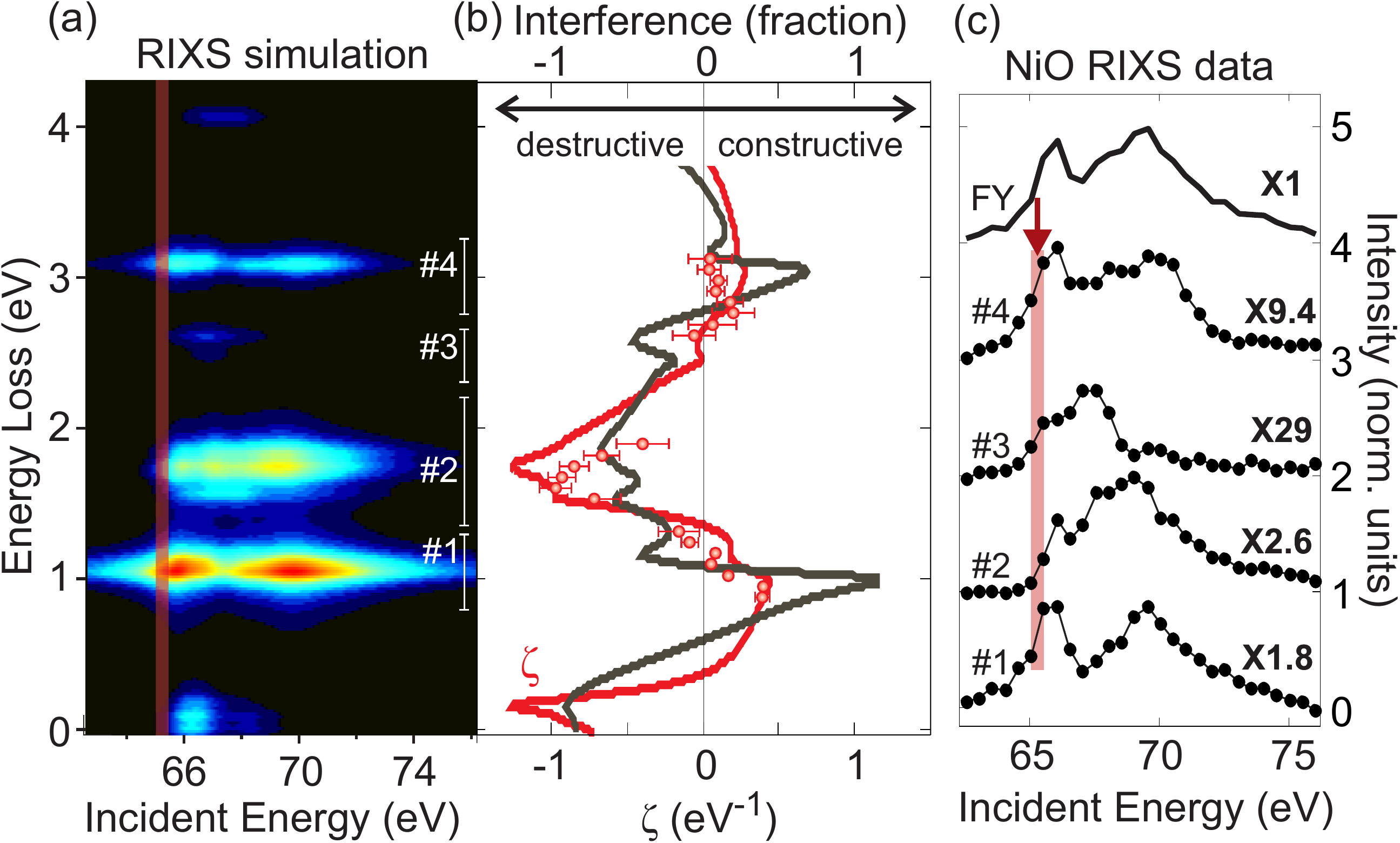}
\caption{{\bf{Quantum interference in the experimental RIXS profile of NiO}}: (a) A simulation of RIXS from NiO is displayed with a red-hot color scale. Energy regions in which intensity is dominated by different inelastic features are numbered 1-4. (b) (red curve) An estimate of quantum interference at $h\nu=65.25eV$ (indicated by a vertical line in panels (a,c)) in the simulated RIXS profile is obtained from the $\zeta(E,h\nu)$ function (Eq. \ref{eq:zetaDef}), and compared with (dark green curve) the \emph{exact} fractional effect of quantum interference on scattering intensity, as defined in the text. (red points) Experimental values of $\zeta(E,h\nu)$ are shown with error bars, where adequate statistics can be obtained. (c), Very different patterns are seen in the incident energy dependence of inelastic fluorescence yield (FY), and RIXS in the four energy loss regions numbered in panel (a). The incident energy at which FY onset has a maximal slope ($h\nu=65.25eV$) is identified with a red stripe.}
\label{fig:zetaFig}
\end{figure*}

The M-edge RIXS spectram of the $3d^8$ Mott insulator NiO is ideal for testing the $\zeta(E,h\nu)$ function, because it has a complicated structure that is nonetheless known to correspond closely with the features in AM simulations \cite{NiO_LRIXS,NiO_MRIXS,WrayNiO} such as that shown in Fig. \ref{fig:zetaFig}(a). The spectrum has four principle features that can be measured experimentally with good statistics (numbered 1-4 on the image). Other notable features that are not observed with adequate statistics to experimentally evaluate the $\zeta(E,h\nu)$ function will be discussed with respect to numerical modeling results, including spin excitations close to the elastic line ($E<0.3eV$) and charge transfer excitations at higher energies ($E\gtrsim5eV$).

Focusing first on the AM simulation results, Fig. \ref{fig:zetaFig}(b) compares an estimate of quantum interference obtained from (red curve) $\zeta(E,h\nu)$ with (dark green curve) an exact calculation of the fractional contribution of quantum interference to the spectrum ($\frac{R(E,h\nu)-R_{off}(E,h\nu)}{R_{off}(E,h\nu)}$). Here, $R_{off}$ is the RIXS intensity when quantum interference has been turned off by moving the sum over $m$ outside of the absolute value brackets in Equation (\ref{eq:KH}).

The $\zeta(E,h\nu)$ function clearly tracks the dark green curve representing the actual contribution of quantum interference. Low energy spin modes found at $E<0.3$eV are correctly identified as having destructive interference, and the prominent $\sim$1eV orbital excitation is identified as having slightly constructive interference. Strong destructive interference in feature $\#$2 is consistent with excitations into the $^3T_1$ branch of the $^3F$ ground state manifold in the $d^8$ Tanabe-Sugano diagram. This feature comes from the simultaneous excitation of two t$_{2g}$ electrons to the e$_g$ level, a process that only becomes allowed due to Coulomb interactions that involve the core hole (i.e. core hole shake-up). Evaluating $\zeta(E,h\nu)$ on the simulation also labels high energy (5-8eV) charge transfer modes as having destructive interference (see Appendix E) which is correct within the numerics, but the corresponding experimental features are too weak to be identified.

Calculating $\zeta(E,h\nu)$ from experimental data (red points in Fig. \ref{fig:zetaFig}(b)) gives reasonably good correspondence with the contour obtained from the simulation for all prominent features in the experimental spectrum. Interference in feature $\#$1 is moderately constructive at the leading edge, but slopes to neutral interference at the trailing edge. This observed pattern confirms the analysis in Ref. \cite{NiO_MRIXS} that the trailing edge of the 1eV feature includes a spinful excitation mode, which must have destructive path-phases at the M-edge (as discussed above for SrCuO$_2$).

\subsection{The cuprate K-edge}

\begin{figure*}[t]
\includegraphics[width = 17cm]{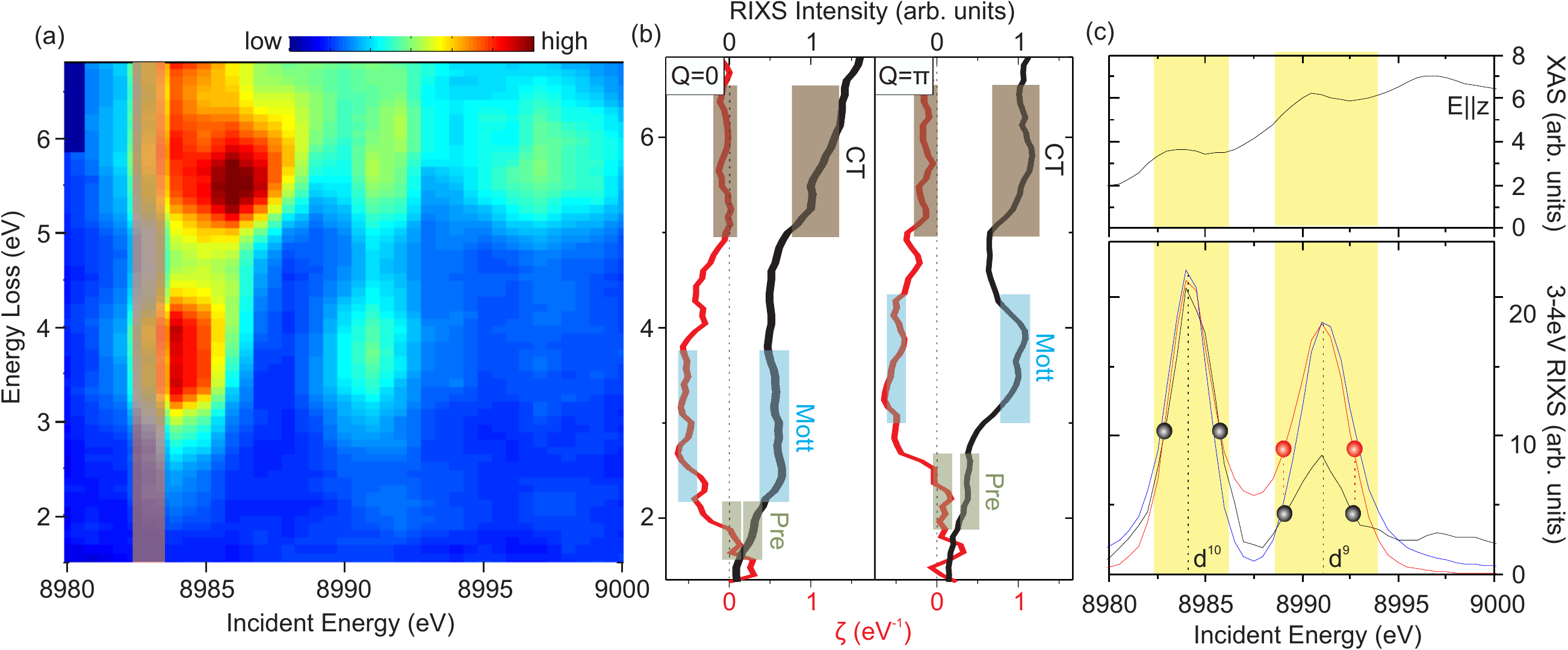}
\caption{{\bf{Quantum interference at the copper K-edge}}: (a) The incident energy dependence of K-edge RIXS is shown for SCO. (b) (black) RIXS curves and (red) the $\zeta(E,h\nu)$ function measured at the leading edge of resonance ($h\nu$=8983eV) are shown for the center and boundary of the 1D single-chain Brillouin zone.  Shaded energy regions are associated with a pre-gap feature, Mott gap excitations, and metal-ligand charge transfer (CT) excitations, respectively. (c) (top) X-ray absorption is estimated from fluorescence yield, with incident polarization matching the RIXS measurements. (bottom, black) Incident energy dependence of RIXS intensity in the 3-4eV Mott gap energy loss window is compared with 2-peak fits that assume (red) fully destructive and (blue) fully constructive path phases. Circles mark points on the red and black curves at which intensity has dropped by 50$\%$ relative to the nearest local maximum.}
\label{fig:Kedge}
\end{figure*}

Measuring quantum interference with deeper-lying core holes gives the valuable opportunity to probe phase information as a function of transferred momentum across the Brillouin zone (BZ). This will provide a way to gauge how an excitation's symmetry changes as a function of momentum, which is particularly important to establish an appropriate model of how different many-body degrees of freedom contribute to dispersion anomalies. Momentum dispersion anomalies (e.g. kinks) will not explicitly impact the evaluation of quantum interference (e.g. via the $\zeta(E,h\nu)$ function), which is performed along the incident energy axis.

The K-edge (1s-4p) RIXS profile of SCO is shown in Fig. \ref{fig:Kedge}(a). The creation of excitations at the K-edge is widely attributed to shake-up processes from the strong monopole potential of the 1s core hole \cite{Hasan_KRIXS,TohyamaSqw,DevereauxSqw,NCCOIshii,AmentRIXSReview}, which splits intermediate states with well screened (d$^{10}$) and poorly screened (d$^9$) electron configuration on the copper site into two resonances at roughly 8984eV and 8991eV (see labels in Fig. \ref{fig:Kedge}(c)). Higher energy resonant scattering observed near 8997eV appears to involve physics beyond a single band Hubbard model (e.g. compared with numerics in Ref. \cite{TohyamaSqw}).  A high resolution energy loss curve ($\delta E \sim 0.1eV$) at the 1D Brillouin zone boundary shows multiple features (Fig. \ref{fig:Kedge}(b)), including Mott gap excitations new 3.5eV \cite{TohyamaSqw,Hasan_KRIXS,HasanSCO_KRIXS,YJKimSCO}, a metal-ligand charge transfer (CT) mode found above 5eV, and a $\sim$2.2eV pre-gap feature that may represent the simultaneous excitation of a Mott gap excitation and two spinons \cite{YJKimSCO}.  The primary Mott gap feature has a large $\sim$1eV dispersion across the Brillouin zone. The pre-gap feature has not been resolved in earlier studies, but was speculated to exist due to an anomaly in the energy loss onset of the Mott gap feature. 

Due to the underlying shake-up mechanism K-edge RIXS, all K-edge excitations are expected to have fully destructive path phase interference, which would cause them to vanish completely if the core hole shake-up potential were set to zero. The red curve in Fig. \ref{fig:Kedge}(c, bottom) shows fits of the Mott gap excitation as a 2-slit fully destructive or fully constructive scattering process, as was done in Section II for SCO orbitons. In this simplified picture, the two scattering paths are through well screened ($d^{10}$) and poorly screened ($d^{9}$) resonance states, and converge on a single final state representing a Mott gap excitation. The fits use the known K-edge lifetime of $\Gamma$=1.55eV and additional gaussian broadening to represent a convolution with the many-body density of states. Gaussian broadening increases from $\sigma$=0.7eV at 8984eV to 1.0eV at $h\nu$=8991.3eV and is constant outside of this range, to represent the growing continuum of allowed resonance states above the resonance edge. The result is a 2-peak spectrum with softer slopes between the peaks for underlying constructive interference, and outside of the peaks for underlying destructive path phases. The asymmetrical experimental line shapes correspond well with the destructive path phase fit, although absolute peak intensities cannot be compared nicely with the fit due to the sharp growth of resonant absorption between the two peaks.

Evaluating the $\zeta(E,h\nu)$ function at the leading edge of resonance (8983eV) shows a consistent pattern at the center and boundary of the 1D BZ (see Fig. \ref{fig:Kedge}(b) left and right panels). According to this metric, the Mott gap feature is excited with the most destructive interference ($\zeta_{Mott}\sim -0.5$), followed by the CT excitation ($\zeta_{CT}\sim-0.1$) and the pre-gap mode ($\zeta_{Pre}\sim 0$). Self absorption is not expected to be a significant source of error here, because the $\zeta(E,h\nu)$ is evaluated at the leading edge of the absorption spectrum, and non-resonant absorption dominates for emitted photons at the principle inelastic features.


The different $\zeta(E,h\nu)$ values associated with each mode can be identified with the time scales of mode creation. As discussed in Section 2A, the fractional intensity of a shake-up excitations in the RIXS spectrum grows roughly as $I\sim K^2/\Gamma^2$ when core hole lifetime is short ($\Gamma \gg K$), before leveling off as core hole lifetime becomes long ($I\sim constant$, as $\Gamma<K$). These regimes represent phase interference in Eq. \ref{eq:KH} that is strongly destructive when formation kinetics are slow relative to the core hole lifetime, and becomes neutral when the kinetics of excitation formation are fast. The Mott gap excitation is created through kinetics set by the intersite Cu-Cu hopping parameter ($t_{CuCu}\sim 0.6eV$ \cite{spinChargeSep}), whereas the CT mode is created via kinetics set by the Cu-O near neighbor hopping parameter (commonly described as $t_{CuO}\sim 2.2eV$ for cuprates). For the Mott gap excitation, the inverse core hole lifetime $\Gamma=1.55eV$ is large relative to $t_{CuCu}$, in keeping with the identification from the $\zeta$ function that it is created via destructively interfering resonance channels.

Intriguingly, the pre-gap feature has a very different $\zeta$ value than the Mott gap feature immediately next to it, strongly supporting the proposal that it represents a distinct excitation symmetry. Ref. \cite{YJKimSCO} identifies this feature via a leading edge anomaly in the RIXS spectrum of the Mott gap feature, starting from roughly 1.7eV in the BZ center and 2.0eV at BZ boundary. These energies match the leading edge of the pre-edge feature in our data to within $\pm0.1eV$. The relatively constructive interference implied by $\zeta_{Pre}$$\sim$0 suggests that this mode forms rapidly, via kinetics that are faster than the separation of a particle-hole pair into delocalized Mott gap excitations. Moreover, the mode has reduced momentum dispersion relative to the Mott gap excitation continuum. Both of these characteristics suggest that the pre-gap mode may have exciton-like character.

\begin{figure}[t]
\includegraphics[width = 8.7cm]{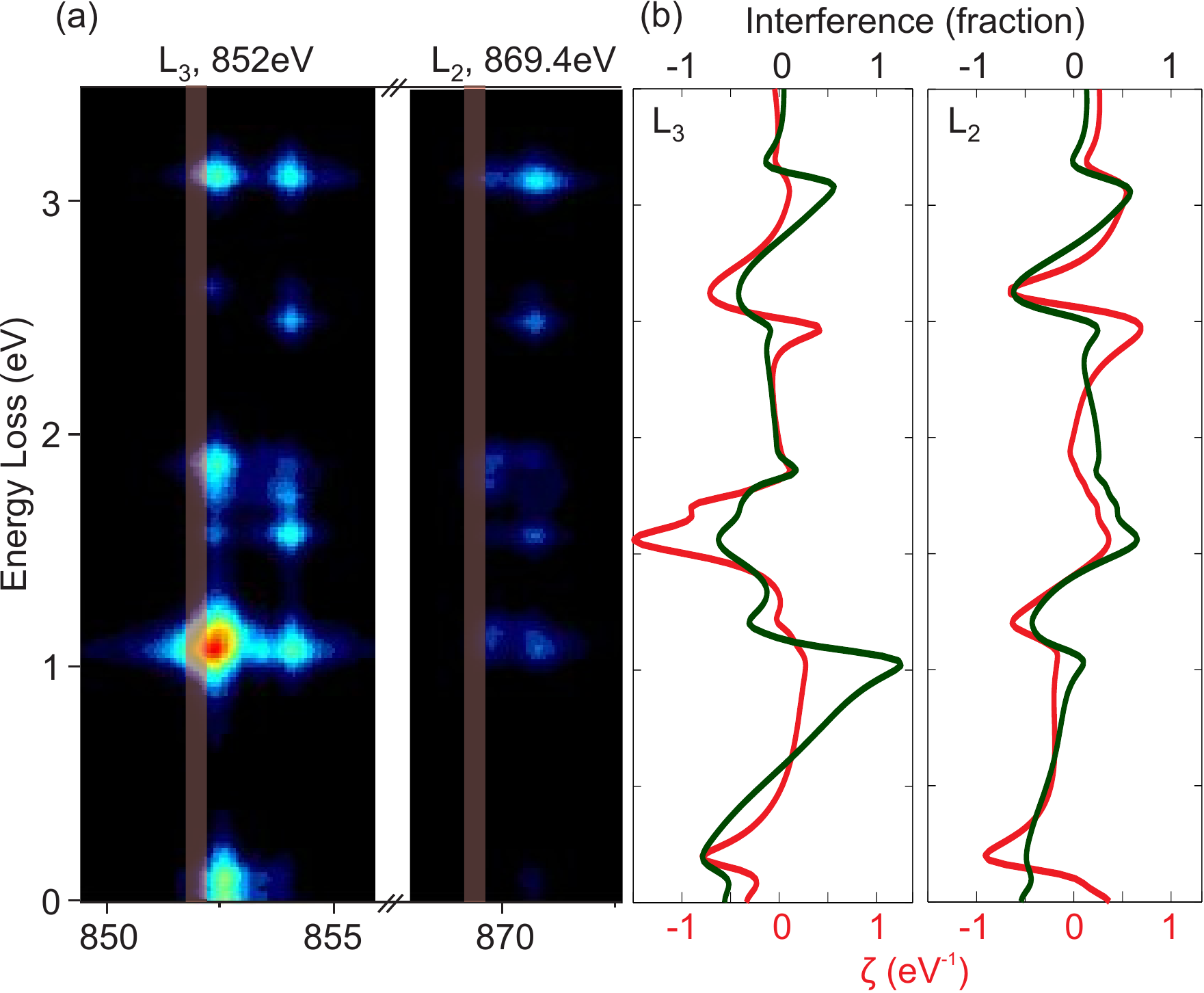}
\caption{{\bf{Predicting interference at the Ni L-edge}}: (a) The incident energy dependence of L-edge RIXS is modeled for NiO. (b) (red curve) Quantum interference at $L_3$ (852eV) and $L_2$ (869.4eV)  is estimated from (red curve) the $\zeta(E,h\nu)$ function (Eq. \ref{eq:zetaDef}), and compared with (dark green curve) the \emph{exact} fractional effect of quantum interference on scattering intensity. Regions at which the $\zeta(E,h\nu)$ curve would be easiest to measure experimentally with good statistics are highlighted in red.}
\label{fig:Ledge}
\end{figure}

\subsection{The L-edge of NiO}


An AM simulation for RIXS at the L-edge of NiO is presented in Fig. \ref{fig:Ledge}(a), and the $\zeta(E,h\nu)$ function at $L_3$- and $L_2$-edges is evaluated based on the simulation in Fig \ref{fig:Ledge}(b). As at the M-edge, there is strong correspondence between the $\zeta(E,h\nu)$ function and the actual fractional contribution of quantum interference, and the line shape of the $\zeta(E,h\nu)$ function has local maxima and minima corresponding to essentially all of the local maxima and minima of quantum interference.

Application of the $\zeta(E,h\nu)$ function to the leading edge of $L_3$ resonance is justified by the derivation in Appendix C, but we note that its application at $L_2$ is not fully justified, to the extent that there is interference between $L_2$ and $L_3$ resonance. This particularly applies to spinful ($\Delta S =\pm 1$) excitations which are classified as photon operator RIXS at the isolated $L_2$- edge \cite{Valentina}, but fit the definition of shake-up RIXS when interference between $L_2$ and $L_3$ is considered. With respect to the definitions in Section 2A, this is established by noting that $\sum_m\langle f|T^{\dagger}| m\rangle \langle m|T|g\rangle \neq 0$ if m indexes only $L_2$ (or only $L_3$) states, but the same sum gives exactly zero when m indexes all states at $L_2$ and $L_3$, and 3d spin orbit coupling is neglected. The principle consequence is that the $\zeta(E,h\nu)$ function cannot be trusted for evaluating the quantum interference underlying creation of spinful modes at $L_2$, and incorrectly associates spinful excitations at 0.1eV energy loss with constructive interference. Lastly, we note that phonon modes may also be distinguishable at the L- and K-edges via the $\zeta(E,h\nu)$ function, and are discussed in Appendix D.

\section{Discussion}

Identifying constructive and destructive scattering phases is particularly important to understand resonant scattering in a time-resolved context. Though the Kramers-Heisenberg scattering equation has explicit time dependence through the $\Gamma_m$ lifetime term in the denominator, it is \emph{implicit} time dependence encoded in path-phases that distinguishes the two principle mechanisms by which excitations are created in RIXS.

\begin{figure}[t]
\includegraphics[width = 8.7cm]{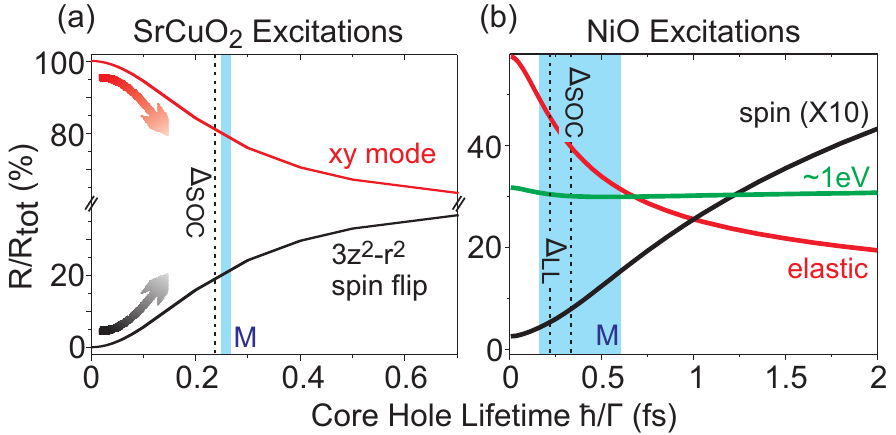}
\caption{{\bf{Transforming the RIXS process into the time domain}}: (a) Fractional intensity of the $3d_{xy}$ orbiton and $3d_{3z^2-r^2}$ spin-and-orbital excitation in the RIXS profile is simulated as a function of core hole lifetime, using Equation (\ref{eq:QIsimpleFit}) with experimental fit parameters. The expected trend for constructive and destructive path-phases is indicated with red and black arrows, respectively. (b) Scattering intensity of several significant final states of NiO is simulated as a function of core hole lifetime using an AM calculation. Curves represent the scattering intensity for (red) elastic scattering, (black) spin rotation excitations near 0.1-0.2eV, and (green) the $\sim$1eV $t_{2g}$ to $e_g$ orbital excitation ($^3T_2$ excitation manifold). Time scales of M-edge core hole lifetime are indicated with blue shaded regions, and dashed vertical lines indicate time scales associated with core hole energetic parameters. The normalizing factor R$_{tot}$ is defined as the sum of $3d_{3z^2-r^2}$ and $3d_{xy}$ excitation intensity for panel (a), and the sum over all atomic multiplet excitations for panel (b).}
\label{fig:SCOtime}
\end{figure}

Destructive path-phase interference is a telltale marker of shake-up excitations \cite{MLdirectNote}, and has the direct physical consequence of suppressing the creation of these modes on short time scales. For example, spin flip excitations such as the 3d$_{3z^2-r^2}$ spin flip orbiton of SrCuO$_2$ and the 0.1-0.2eV modes of NiO are created with destructive path-phase interference, and are therefor more likely to be created if core hole lifetime is long. The manifestation of similar physics in magnetic excitations at the transition metal L-edge is discussed in Ref. \cite{Valentina}. This observation can be made quantitative by using the Kramers-Heisenberg equation to plot a time-dependent picture of how the intermediate state projects onto final state excitation symmetries, using scattering path parameters obtained from experiment for SrCuO$_2$ (Fig. \ref{fig:SCOtime}(a)) and from theory for NiO (Fig. \ref{fig:SCOtime}(b)). The $\zeta(E,h\nu)$ function, by contrast, gives more qualitative information.

Excited states that are directly accessed by the action of the incident and outgoing photon operators are not suppressed like shake-up excitations when core hole lifetime is very short. When path-phase interference is close to neutral, as identified by the $\zeta(E,h\nu)$ function for the 1eV NiO excitation, there is very little intensity dependence on core hole lifetime. Scattering paths that return to the ground state (i.e. elastic scattering) have strongly constructive path-phase interference, and are suppressed on longer time scales due to the increasing probability of shake-up excitations (Fig. \ref{fig:SCOtime}(b)). The sub-femtosecond time evolution is determined by energetic interactions such as core hole spin-orbit coupling (SOC) and angular momentum coupling between the core hole and valence electrons (LL coupling, $\Delta_{LL}=E(^3D)-E(^3F)\sim 4eV$ for NiO), labeled with dashed lines in Fig. \ref{fig:SCOtime}.

Taken collectively, the experimental results presented here give a rough picture of what quantum phase interference looks like in inelastic scattering at the three most commonly accessed transition metal resonance edges, and how it can be used to distinguish between different classes of excitation. Constructive interference between scattering channels is associated with resonance intensity that is more spread out on the incident energy axis, and destructive interference is associated with resonance pinched into a narrower energy band. From the $\zeta(E,h\nu)$ function, we see that constructive interference generates a shallower slope at the leading edge of spectra. These patterns are opposite to what is seen in real-space elastic scattering, where constructive interference between domains is associated with \emph{sharper and narrower} scattering features. This discussion is fundamental to widely observed scattering effects \cite{AmentRIXSReview,VDBultrashort,DevereauxSqw,Valentina,FC_K}, but differs from earlier experimental studies in that it does not depend on any detailed model of the intersite many-body state. When X-ray resonances are broadly separated relative to their inverse lifetimes, these interference patterns will evolve into Fano lineshapes \cite{Fano}, such as have been experimentally observed in the lanthanum ($4f^0$) $N_4$ and $N_5$ RIXS spectrum \cite{RIXSfano}. In this limit, destructive (constructive) path-phase interference can be identified from Fano tails pointing towards (away from) the centroid of resonance. For example, the destructive interference we have observed between $M_2$ and $M_3$ resonances for spin-rotation excitations will manifest as inward-pointing Fano tails between the $L_2$ and $L_3$ edges (e.g. for single magnon modes).

Measurement of quantum interference may in some cases be complementary or similar to the measurement of scattered photon polarization.  Both techniques can be used to distinguish between spinful and spineless modes, and the ``photon operator" or ``shake-up" nature of of quantum interference (see definitions in Section 2A) can also be a key factor in polarization analysis. This is because photon operator RIXS is by definition linked to the incident and outgoing polarization vectors, while shake-up RIXS is often a spectator process in which the symmetry of the final state RIXS excitation does not directly determine scattered photon polarization. A comparative advantage of quantum interference measurements via the $\zeta(E,h\nu)$ function is that they do not require additional instrumentation or a much longer measurement time than standard RIXS.

In conclusion, the results in this paper have shown that quantum interference provides a spectral fingerprint with important information for evaluating the nature of low energy RIXS excitations. This can be done in a model independent way, by associating destructive interference with slower shake-up excitations, or with respect to specific numerical predictions. At the transition metal M-edge extremely good energy resolution can be achieved, and measuring quantum interference will give a new dimension of information for identifying the symmetries of newly resolved excitations. At higher energy resonance edges (L- and K-), the analytic methods presented here work synergistically with momentum resolution, and are used to identify the presence of a novel excitation mode a the edge of the SCO Mott gap, which is only resolvable at momenta far from the Brillouin zone center. As upcoming spectrometers overcome current data quality limitations, there will be ample room to develop improved metrics of quantum interference throughout the RIXS profile. In the further future, stimulated RIXS experiments may make it trivial to identify quantum interference by enabling control of the core hole lifetime \cite{stimRIXS}.

\section{Acknowledgements}

We are grateful for discussions with K. Wohlfeld, B. Moritz, T. Devereaux, K. Ishii, I. Jarrige, R. Eder, D.-H. Lee, S. Roy, and P. Shafer. M.Z.H. is supported by U.S.DOE/BES grant number DE-FG02-05ER46200. The Advanced Light Source is supported by the Director, Office of Science, Office of Basic Energy Sciences, of the U.S. Department of Energy under Contract No. DE-AC02-05CH11231. This research used resources of the Advanced Photon Source, a U.S. Department of Energy (DOE) Office of Science User Facility operated for the DOE Office of Science by Argonne National Laboratory under Contract No. DE-AC02-06CH11357.

\setcounter{equation}{0} \renewcommand{\theequation}{A\arabic{equation}}
\renewcommand{\thetable}{A\arabic{table}}

\setcounter{figure}{0} \renewcommand{\thefigure}{A\arabic{figure}}

\section{Appendix}

\subsection{Experimental methods}

Measurements at the M-edges of Ni and Cu were performed at the beamline 4.0.3 (MERLIN) RIXS endstation (MERIXS) at the Advanced Light Source (ALS), Lawrence Berkeley National Laboratory, and measurements at the K-edge of Cu were performed at the Advanced Photon Source Sector 30 MERIX beamline. Large single crystal samples were measured near room temperature at a pressure of 3$\times$10$^{-10}$ Torr. The resolution-limited full width at half maximum of the elastic line is better than $\delta$E$\lesssim$35$\pm$2 meV for all M-edge RIXS measurements, and $\delta$E$\lesssim$100 meV at the K-edge. Inelastic fluorescence yield is obtained from integrating the intensity of all excitation features observed by RIXS, and has been used as $I_{XAS}(h\nu)$ in the $\zeta(E,h\nu)$ function. All M-edge measurements were performed using in-plane ($\pi$) polarization in the [001] scattering plane, with scattered photons measured at 90$^o$ to the incident beam trajectory, and incident photons at a grazing 25$^o$ angle to the cleaved [010] face. The in-plaquette Cu-O axes x- and y- correspond to the [001] and [010] directions respectively, in the crystallographic notation for SrCuO$_2$. Measurements at the copper K-edge were performed with $\sigma$ polarization, with the polarization axis normal to the Cu-O plaquettes and momentum transfer along the chain direction.

Statistical error in M-edge measurements arises mostly from CCD dark noise. In Fig. 3, jitter in the background between scans is minimized by smoothing $\zeta(E,h\nu)$ values over a $\pm$0.5eV incident energy range. Systematic error from the discrete derivative is expected to slightly reduce the amplitude of all experimental $\zeta(E,h\nu)$ values.

\subsection{Computational details}

Atomic multiplet calculations use a 3p spin orbit coupling strength of (NiO) 1.6eV and (SrCuO$_2$) 1.8eV. For SrCuO$_2$, d-orbital spin orbit coupling is disregarded, leaving only crystal field energetics in the d-orbital Hamiltonian (reflected in the orbital energies). AM calculations for Fig. 3 of the main text use atomic values (80$\%$ of Hartree-Fock) for inter-d-orbital Slater-Condon interaction parameters, and are augmented by a single impurity Anderson model (SIAM). Core-valence 3p-3d interactions ($G1(pd)$,$G3(pd)$,$F2(pd)$) require a larger correction at the M-edge relative to the L-edge, and are estimated at 80$\%$ of nominal atomic values. Parameters of the SIAM model mostly follow the treatment in Ref. \cite{KotaniSIAM}. We define a ligand band with W=3eV bandwidth, similar to the maximum dispersion observed for an oxygen derived band in recent analysis of NiO \cite{ShenNiO,EderBandWidth}. Hopping between $e_g$ orbitals and the nearest neighbor ligand states with corresponding symmetry is $V_{eg}$=2.2eV ($V_{t2g}=-V_{eg}/2$), and is reduced by 10$\%$ when a core hole is present. The difference between average configuration energies for $d^{n}$ and $d^{n+1}$ electron occupancy is set to $\Delta$=3.5eV for the ground state, and modified by a core hole potential of $U_C$=7.6eV, and d-orbital Hubbard U parameter of $U_d$=7.2eV in the intermediate resonance states.  The state basis is limited to allow a maximum of 1 ligand hole, as in Ref. \cite{MagnusonSIAM}, and the flat ligand band density of states distribution is approximated by 10 discrete energies (N=10). The ground state density of oxygen holes within this model matches the experimentally based estimate of 0.2 \cite{NiOp2holes}. The AM implementation in Fig. 6 of the main text does not include the SIAM ligand band, so as to focus on atomic multiplet excitations, and uses a correspondingly larger crystal field parameter of 10Dq=1.03eV.

Interatomic spin interactions are considered by incorporating an external exchange field of $J^*$=0.1eV for NiO. The quasi-1D spin lattice and large spin interactions of SrCuO$_2$ result in corrections to the self energy contour of orbitons \cite{Jeroen1Dfluct,spinOrbitalSep}, but do not behave as a static exchange field for most cuprate orbiton symmetries \cite{SalaLedge}. This model and the method used to fit quantum interference in SrCuO$_2$ apply when all intersite interactions in the intermediate state Hamiltonian (such as e.g. spin exchange J) are much smaller than the inverse core hole lifetime, and can be neglected. Though the ground state of SrCuO$_2$ is treated as having no spin polarization, the spin matrix elements for creating the $3d_{3z^2-r^2}$ orbiton are dominated by Pauli matrices aligned in the x-z plane, and one must choose a spin quantization axis normal to this Pauli vector (for example, the y-axis) to frame the orbiton as occurring with a spin flip. When exciting the $3d_{xy}$ orbiton, spin is primarily acted on by the identity matrix, and no choice of quantization axis gives more than 3$\%$ spin flip intensity (averaged over the incident energy axis).

\subsection{Quantum interference and the onset line shape of resonance}

The following equation based on Eq. 2 in the main text provides a simple starting point to look at the effect of quantum interference on line shapes at the leading edge of core hole resonance:

\begin{align} \label{eq:leadingEdgeKH}
R_f(h\nu)=\left| A_{f,1}G_{1}(h\nu)+B_f \right|^2
\end{align}


Here, the RIXS intensity of a given final state ($R_f$) is evaluated at an energy $h\nu$ that is slightly lower than the first core hole resonance state, indexed by m=1 (e.g. this would have $M_3$ symmetry for SrCuO$_2$). The real valued amplitude ($A_{f,m}$) and Green's function ($G_m(h\nu)=(h\nu-E_m+i\Gamma_m/2)^{-1}$) of the lowest energy core hole state are considered explicitly, while the tail of resultant quantum amplitude from all higher energy states is approximated by a real valued parameter (B$_f$) that is negative for constructively interfering path-phases and positive for destructive phases. The effect of quantum interference on the onset line shape of resonance can be summarized by taking an intensity-normalized derivative:

\begin{align}
\zeta_f(h\nu)=-R_f(h\nu)^{-1} \frac{\mathrm{d} R_f(h\nu)}{\mathrm{d} h\nu}
\end{align}

This quantity `$\zeta_f$' has the useful property that all RIXS excitations created with net constructive path-phases have larger $\zeta_f$ values than destructive path-phase RIXS excitations, when considering the simplified scattering equation in Equation (\ref{eq:leadingEdgeKH}). The main text defines an analogous quantity that can be evaluated from experimental RIXS spectra:

\begin{align} \label{eq:zetaDef}
\zeta(E,h\nu)=I(h\nu)^{-1} \frac{\mathrm{d} I(h\nu)}{\mathrm{d} h\nu}-R(E,h\nu)^{-1} \frac{\mathrm{d} R(E,h\nu)}{\mathrm{d} h\nu}
\end{align}

Here, $R(E,h\nu)$ is the RIXS intensity at incident energy $E$ and energy loss $h\nu$, and X-ray absorbtion ($I(h\nu)$) is used to center the function on zero. Good statistics are obtained by evaluating $\zeta(E,h\nu)$ at an incident energy that has maximal slope in the XAS profile.


\subsection{Franck-Condon phonon shake-up \label{sec:PhononAppendix}} 

Deeper-lying core holes are known to cause the entire atomic wavefunction of the scattering site to contract, leading to strong Franck-Condon phonon shake-up and other many-body shake-up features. These many-body effects will shifts peak energies, much like a the magnon shake-up effect creates incident energy dependence in the energy loss centroid of the RIXS 1eV feature in NiO. In some cases, when shake-up is associated with a degree of freedom that one is not interested in studying, it may also be useful to disregard energy loss shifts by applying the $\zeta(E,h\nu)$ function to the energy-loss-integrated intensity of a feature.

Multi-phonon shake-up has been associated with incident energy dependent shifts of $<$10meV \cite{WrayCoO}, $\sim$50meV \cite{FC_L} and $>$0.1eV \cite{FC_K} in excitation energies at the M-, L-, and K-edges of 3d transition metals, respectively. Hartree-Fock calculations can qualitatively explain the degree of shake-up occurring at each edge, as they predict that the radial wavefunction of d-electrons in NiO will contract by (M) 0.6$\%$, (L) 6$\%$ and (K) 11$\%$ when core holes of these symmetries are present. This means that quantum interference features are easier to relate to a purely electronic model at the M-edge, but will reflect a broader range of physics at other edges. In the case of the Franck-Condon effect, data and calculations show that the intermediate state induced shake-up component of excitations vanishes below the leading edge of resonance. This pattern can be understood from the discussion of leading edge intensity in Appendix C. The $\zeta(E,h\nu)$ function can be used to help identify the Franck-Condon-shifted components of excitations, as it will highlight them as having destructive path-phase interference.

\subsection{Using quantum interference to learn about the nature of excitations: photon operator RIXS and shake-up RIXS}


Identifying ``photon operator" RIXS and ``shake-up" RIXS scattering mechanisms (defined in Section 2A) can provide a good starting point for anticipating the nature of quantum interference in a RIXS spectrum. Generally speaking, excitations created by a shake-up processes will have strongly destructive path-phase interference compared to excitations that occur through photon operator RIXS. We note that ``photon operator" and ``shake-up" RIXS are called ``direct" and ``indirect" RIXS respectively in Ref. \cite{AmentRIXSReview}, but we have avoided this terminology as it is sometimes inconsistent with recent usage of ``direct RIXS" to indicate resonant scattering that involves direct transitions between the core and valence levels (for example, transition metal L-edge RIXS).

Table \ref{tab:intensityME} shows the scattering process principally associated with different types of RIXS excitations. In many cases, the identification with photon operator or shake-up RIXS is a nuanced question. Excitations that are clearly shake-up RIXS in the limit of slow core-hole-relate dynamics (e.g. single magnons and most Mott gap excitations) will behave as photon operator RIXS when the relevant core hole energetic parameter is very large relative to the inverse core hole lifetime. The key parameter deciding this for single magnons is core hole spin-orbit coupling, and the most significant parameters for Mott gap excitations are the core hole monopole potential and intersite hopping parameters.

\begin{table}[h]
\centering
\begin{tabular}{lccccc}
Edge: &      M & $L_3$  & K$_{pre}$ & K$_{WS}$ \\ \hline
\multicolumn{1}{l|}{Phonon}&       shake-up & shake-up & shake-up & shake-up \\
\multicolumn{1}{l|}{Spin rotation}&       shake-up & PO & N/A & N/A \\
\multicolumn{1}{l|}{Multimagnon shake-up} &      shake-up & shake-up & shake-up & shake-up \\
\multicolumn{1}{l|}{Single-atom excitons} &     PO & PO & PO & shake-up \\
\multicolumn{1}{l|}{Mott gap excitations} &      both & shake-up & shake-up & PO$^*$ \\
\end{tabular}
\caption{\textbf{Associated scattering process in transition metal oxides}: Scattering mechanisms associated with different excitation types are indicated (``shake-up" or ``photon operator" (PO)). The `Spin rotation' heading includes cuprate single paramagnon excitations. The K$_{pre}$ and K$_{WS}$ columns refer to energetically isolated pre-edge (quadrupole excitation) and well screened K-symmetry resonance states, respectively. $^*$\emph{Interatomic exciton modes will likely be more photon operator-RIXS-like than true continuum Mott gap excitations, because the kinetics by which particles separate into a continuum are time-delayed}.}
\label{tab:intensityME}
\end{table}

One can loosely associate M- and L- edge atomic multiplet RIXS features with photon operator RIXS, because the core hole monopole ``shake-up" potential does not factor into simple atomic multiplet models. In practice however, even the simplest models of M- and L-edge scattering typically include a significant contribution from shake-up RIXS, due to various types of angular momentum coupling that involve the core hole and photoexcited electron.

An important aspect of this classification scheme is that the definitions depend on how one chooses to define the intermediate resonance state basis. For example, magnetic excitations that alter the total spin of the system ($\delta$S=1) are allowed via photon operator RIXS if the intermediate state basis includes only $J=3/2$ core hole states (e.g. $L_3$ and $M_3$), but become disallowed when the complimentary $J=1/2$ resonance manifold ($L_2$ and $M_2$) is also included in the RIXS calculation. At the L-edge when the separation between the destructively interfering $J=3/2$ and $J=1/2$ core hole states is large relative to inverse core hole lifetimes ($\Gamma_m$), it is appropriate to say that the $L_3$ and $L_2$ resonance manifolds are accessed separately, and $\delta$S=1 excitations occur via photon operator RIXS. Conversely, at the M-edge, splitting between $J=3/2$ and $J=1/2$ states occurs on a similar energy scale to the inverse lifetime, and $\delta$S=1 magnetic excitations unambiguously occur through \emph{shake-up} RIXS. The same distinction can apply to Mott gap excitations at the K-edge: when the energy separation between well screened and poorly screened resonance states is large, Mott gap excitations will appear to occur via photon operator RIXS.

\begin{figure}
\centering
\includegraphics[width = 8.7cm]{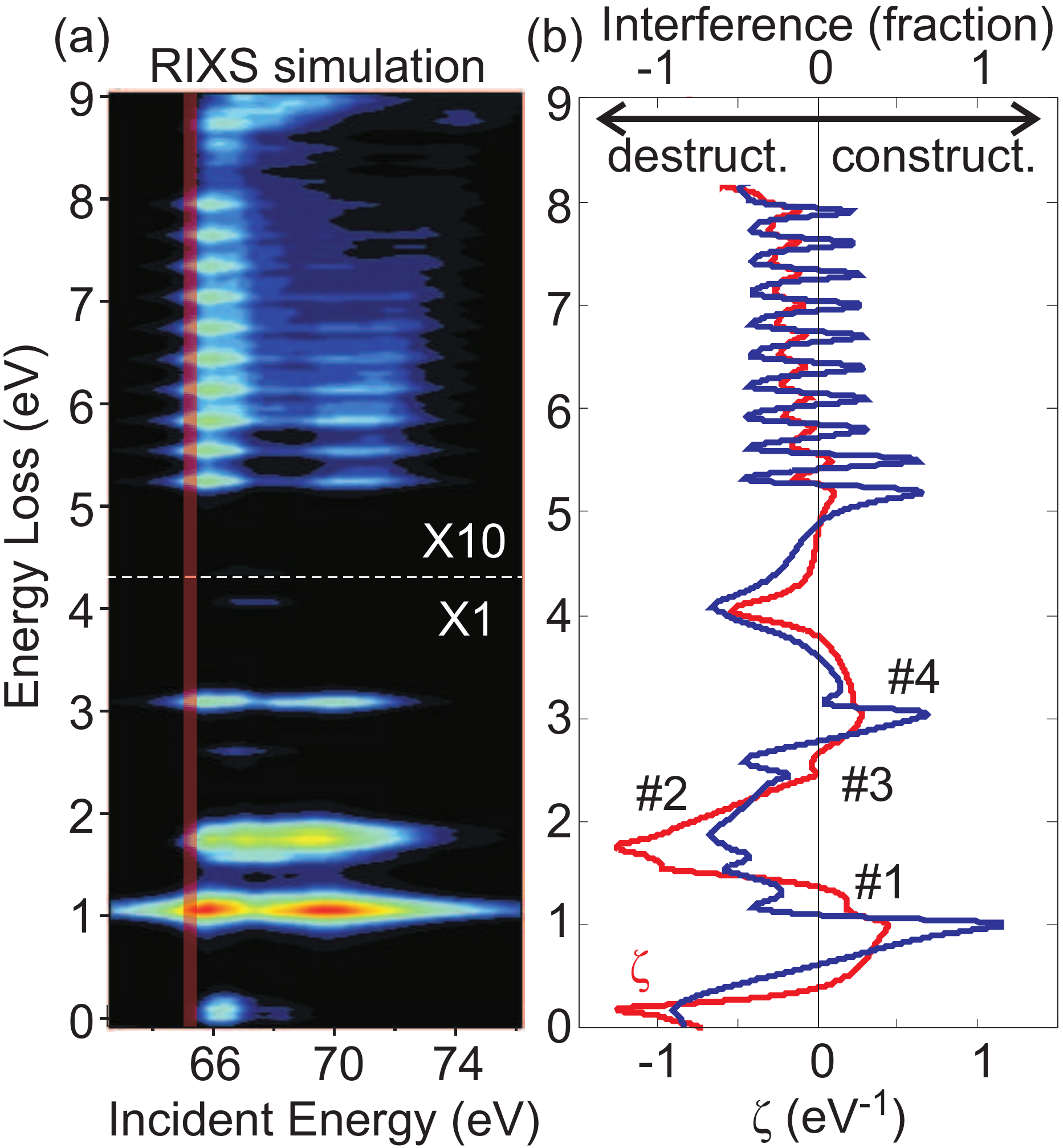}
\caption{{\bf{Quantum interference above the band gap}}: (a) A simulation of RIXS from NiO is displayed with a red-hot color scale. Intensity above the insulating band gap is multiplied by 10 for better visibility. (b) (red curve) The $\zeta(E,h\nu)$ function defined in the main text is used to estimate quantum interference at $h\nu=65.25eV$ in the calculated RIXS profile. (dark blue curve) The exact fractional amplitude of quantum interference is evaluated from the AM calculation, as described in the main text.}
\label{fig:zetaFig2}
\end{figure}

Figure \ref{fig:zetaFig2} shows a version of Fig. 3 from the main text that has been expanded to include charge transfer excitations seen at $\sim$5-8eV in the SIAM model. The rippled structure of these modes is an artifact from numerical discretization of the oxygen ligand band. Although Mott gap excitations are widely associated with shake-up RIXS, they also have a non-zero photon operator RIXS matrix element. At the M-edge, the shake-up RIXS matrix element for these excitations is particularly weak, and the contribution from photon operator RIXS can not be fully neglected.

\begin{thebibliography}
\bibitem{MLdirectNote} Ref. \cite{AmentRIXSReview} divides resonant scattering into ``direct" and ``indirect" classes, with ``indirect" scattering representing shake-up processes which by construction have 100$\%$ destructive path-phase interference. This definition is sometimes inconsistent with recent usage of ``direct RIXS" to indicate resonant scattering that involves direct transitions between the core and valence levels (e.g. L-edge RIXS).
\end{thebibliography}
\end{document}